\documentclass[11pt]{article}
\usepackage{cospar}
\usepackage{psfig}

\begin{document}
\title{IN SITU MEASUREMENTS OF THE PHOBOS MAGNETIC FIELD DURING THE PHOBOS-2 MISSION}
\author{{V.G. Mordovskaya$^{1}$ and V.N. Oraevsky}\address{Institute of
Terrestrial Magnetism, Ionosphere, and Radiowave Propagation (IZMIRAN),
Troitsk, Moscow region, 142190 Russia, mail: valen@izmiran.rssi.ru}}

\maketitle

\begin{abstract}
In this communication, we examine in situ observations of the magnetic
field in the vicinity of Phobos on the {\it Phobos-2} mission and give
some analysis of the data during a unique experiment ``Celestial
Mechanics,'' which leads to the support of evidence of the Phobos
magnetic field. In particular, it is suggested that the peculiarity of
the solar wind interaction with Phobos and rotating direction of the
magnetic field obtained on the circular orbits around Mars are the
evidence for the existence of an intrinsic planetary field of Phobos.
\end{abstract} 

\section*{INTRODUCTION}

The principal objective of the {\it Phobos} mission was to investigate
Phobos, its substance, and the solar wind interaction with Phobos. With
this aim the project had an experiment ``Celestial Mechanics''
~(Kolyuka et al., 1991). Unfortunately, the spacecrafts were lost.
However, {\it Phobos~2} spacecraft was lost when it was in the vicinity
of Phobos. Therefore, some part of the experiment was carried out in
the Phobos vicinity but the data obtained were not studied, moreover,
most of the data are unpublished. In the present paper, we study the
measurements of the magnetic field in the vicinity of Phobos.  The
 disturbances was observed on the circular orbits near Phobos at
distances of 180--250~km from its center when Phobos was in a solar
wind.  During these events, a sharp rise in the regular part of the
magnetic field was observed. Analyzing the interaction of the obstacle
with the solar wind plasma, Mordovskaya et al. (2001) came to the
conclusion, that Phobos has its own magnetic field. Using the equation
of pressure balance for the solar wind and the magnetic field of Phobos
at the magnetopause, in the dipole approximation the estimate of the
magnetic moment of Phobos $M'$ was obtained, $M'\simeq10^{15}$
A$\cdot$m$^2$.

According to the measurements performed in the solar wind, the
projection of the Phobos magnetic moment onto the Mars ecliptic plane
points in the direction of the trailing side of Phobos, i.e., to the
opposite direction of its motion along the orbit around Mars. Phobos
represents itself as a magnetized body. Phobos rotates around Mars,
turning to it by the same side. It always presents the same face to a
hypothetical observer on Mars. This peculiarity of the rotation of the
magnetized Phobos results in the magnetic field signatures, which,
specifically the direction, are phase locked with Phobos rotation rate.
The direction of the magnetic field observed should correspond to the
direction given by the rotation. To resolve this issue completely, in
situ measurements in the Phobos vicinity are required. Unambiguous
evidence of the Phobos magnetic field will only be obtained from the
direct observations. This problem will be explored in the present
paper.

\section*{THE NAVIGATION ASPECTS OF THE PHOBOS MISSION}

The {\it Phobos} space project had a very complicated scheme of flight
relative to Phobos and the organization of a unique experiment named
``Celestial Mechanics.'' A description of the spacecraft flight profile
is given by Kolyuka et al. (1991). The spacecraft was transferred onto
such an orbit around Mars on March 22, 1989 and it remained
permanently within a vicinity of Phobos until March 27. The
semi-major axis and plane of the spacecraft orbit coincided with those
of Phobos and there is a small difference in the orbital
eccentricities. Then, the motion of the spacecraft relative to Phobos
takes the form of ellipses around Phobos, hence, the name
quasi-synchronous orbit.

\vskip 3mm
\begin{minipage}{85mm}
\rightskip 5mm
\psfig {file=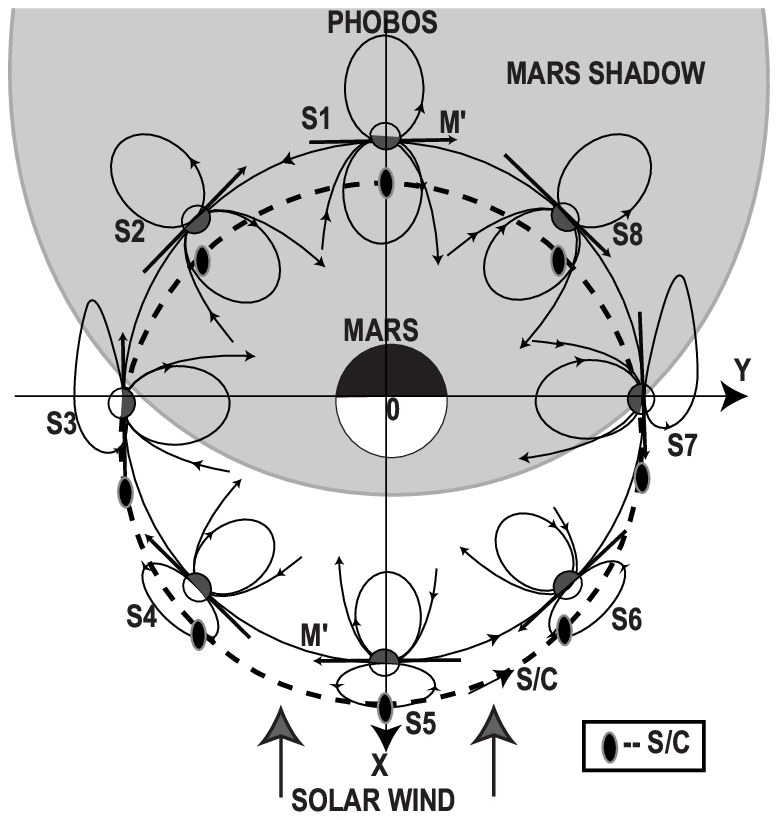} 
{\small {\bf Fig. 1.}
{\bf T}he sketch of the measurements during the
``Celestial Mechanics'' experiment on March 22--26, 1989 and a
schematic model of the Phobos magnetic field with its rotation around
Mars are presented in the projection onto the Mars ecliptic plane
$XOY$.}
\end{minipage}
\begin{minipage}{90mm}
\vspace{-8mm}
Depending on the position of the spacecraft relative to Phobos, the
spacecraft either increases or reduces the orbital velocity but does
not move far away. This circumstance will be important in our analysis.
Figure~1 shows the trajectory projection of the relative motion of {\it
Phobos~2} (S/C, the dashed line) and Phobos (the solid line) onto the
Mars equatorial plane $(XOY)$. The $X$-axis points to the Sun; the
$X$-$Y$ plane coincides with the orbital plane of Mars; the $Y$-axis
points in opposite direction of the Mars's orbital velocity; the
$Z$-axis is perpendicular to $X$ and $Y$.  Thereby defined coordinate
system will be used in the following. The darkened region is the Mars
wake. The dashed line denotes the S/C orbit marked by the S/C image.
The solid line is the Phobos orbit marked by the Phobos image with its
magnetic moment and field. The parameters S1, S2, S3, S4, S5, S6, S7,
and S8 characterize the different situations and distances between the
spacecraft and Phobos on these orbits. At the positions S1 and S5, the
S/C and Phobos are 180--200~km apart and located on the straight line
Mars--Sun. At the positions S2, S4, S6, and S8 the distances between
them are 200--300~km. At the positions S3 and S7 they are 300--400~km
apart.
\end{minipage}

\vskip 10mm
\section*{THE MAGNETIC FIELD MEASUREMENTS NEAR PHOBOS \\
ON THE QUASI-SYNCHRONOUS ORBIT}

Onboard the spacecraft, the two three-axes fluxgate sensors of the FGMM
and MAGMA instruments carried out the magnetic field measurements.  A
detailed description of the magnetometers and the first results can be
found in ~(Auster et al. (1990), Aydogar et al. (1989), M\"{o}hlmann et
al. (1990), Riedler et al. (1989)). We will examine 45-s resolution
magnetic field measurements on circular orbits from March 22 until
March 26, 1989 to present the experimental evidences of the Phobos
magnetic field.

\subsection*{The Solar Wind Interaction with Phobos}

The main argument of the evidence of an intrinsic magnetic field of
Phobos is peculiarities of the solar wind interaction with Phobos.
First, the effective scale of the Phobos obstacle for the solar wind
was observed. Phobos, due to its intrinsic magnetic field forms a large
obstacle, which makes the solar wind be shielded from the surface,
leading a region of about 170 km from dayside surface (the
sizes of Phobos are $18\times21\times27$ km). Second, the subsolar
stand-off distance of the deflection depends on the solar wind
parameters. The morphology of magnetic field signatures due to the
interaction of the Phobos with the solar wind plasma during the time
interval of March 22--26, 1989 was studied by Mordovskaya et al. (2002).
Here we consider the evidence of the existence of the Phobos
magnetic field using data obtained from 18:00 to 20:00 on 24 March,
1989 when we can see the pure planetary magnetic field. The magnetic
field signature is displayed during this encounter with a planetary
magnetic field of Phobos in Fig.~2a. In the middle pannel of the
Fig.~2 the plasma parameters are given. During this period when the
spacecraft approached to Phobos at a distance of 170--180 km from its
dayside the magnetic field signature has a sharp rise with a
characteristic ``shock-like'' behavior, which is marked by the arrows.
The velocity V and density n obtained demonstrate a lack of the solar
wind plasma. This fact gives a strong argument in favour of the
existence of the Phobos magnetic field and its magnetosphere.
\vskip 8mm
\psfig {file=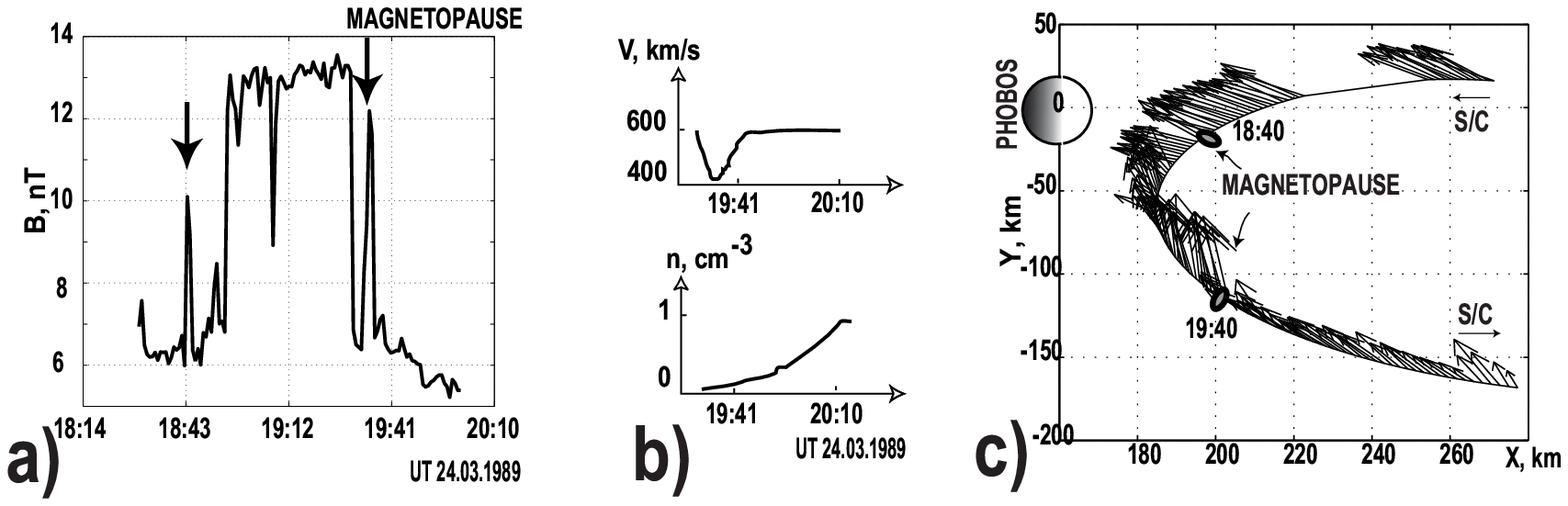}

{\small {\bf Fig. 2.} {\bf a)} The plot of the magnitude of the
observed magnetic field ({\bf B}) versus the time. {\bf b)} The plot
of the velocity V and the density n of the plasma; {\bf c)}
The amplitudes, the direction of the vectors of the magnetic field
Bx/By and the satellite (S/C) trajectory are given in the Phobos centric
coordinate system for 18:00--20:00 on 24 March, 1989.}
\vskip 8mm
The value of the magnetic moment $M'$, which has been calculated with
the use of the equation of pressure balance for the solar wind and
dipole field of the planet on the magnetopause, is $M'\simeq10^{15}$
A$\cdot$m$^2$.

The {\it Phobos-2} encounter with a planetary magnetic field of Phobos
described by help of magnetic field lines is shown in Fig.~2c). The
amplitudes, the direction of vectors of the magnetic field {\bf Bx/By},
and the satellite trajectory (S/C) are given in the Phobos centric
coordinate system. The field observed along the trajectory at 45-s
intervals is represented by a scaled vector projection of {\bf B}
originating from the position of the spacecraft at the
corresponding times. Analyzing the magnetic field direction near Phobos
we can determine with a great precision the boundary between completely
open field lines of the solar wind and those with at least one end in
Phobos. A dramatic change in field line topology from 18:43 to 19:41 on
March 24, 1989 is an evidence indicating a transition from field lines
with no connection to Phobos to field lines with at least one end in
Phobos.

\subsection*{The Corotating Part in the Measured
Magnetic Fields}

Other evidence for an intrinsic magnetic field is the existence of the
corotating part in the measured magnetic fields near Phobos. To
demonstrate the rotation of the direction of the magnetic fields we
will employ the method describing the magnetic events with help of the
magnetic force lines.

According to the peculiarity of the rotation of Phobos around Mars, the
magnetic moment is directed in opposite sides in the Mars wake and
solar wind. Figure~1 depicts the rotation of the magnetized Phobos
around Mars in the projection onto the Mars ecliptic plane.  Comparing
the data, which were obtained and predicted from the peculiarity of the
rotation of Phobos around Mars, it is necessary to remind the
spacecraft was located permanently in the vicinity of Phobos at this
time and the magnetic field data obtained should indicate the rotation
of the direction of the Phobos magnetic field. In addition, one should
take into consideration that, depending on mutual position of the S/C
and Phobos, there exist different situations like S2, S4, S6, and S8.
The magnetometers acquired the vector measurements of the Phobos
magnetic field beginning at various part of the Phobos vicinity.  There
exist the following four situations: the S/C moves from Phobos (S2);
Phobos approaches the S/C (S4); Phobos moves from the S/C (S6); the S/C
approaches Phobos (S8).

Three typical examples of the data measured continuously are
represented in Fig.~3. The field observed along the trajectory at
45-s intervals is represented by a scaled vector projection of {\bf B}
originating from the position of the spacecraft at such times. In this
view, the Mars shadow is marked by the dark line. The spacecraft
trajectory is marked by encounter times of year 1989,
hour:minute~day.month~UT. The corresponding locations of Phobos at
these moments can be found in the scheme shown in Fig.~1.

\psfig {file=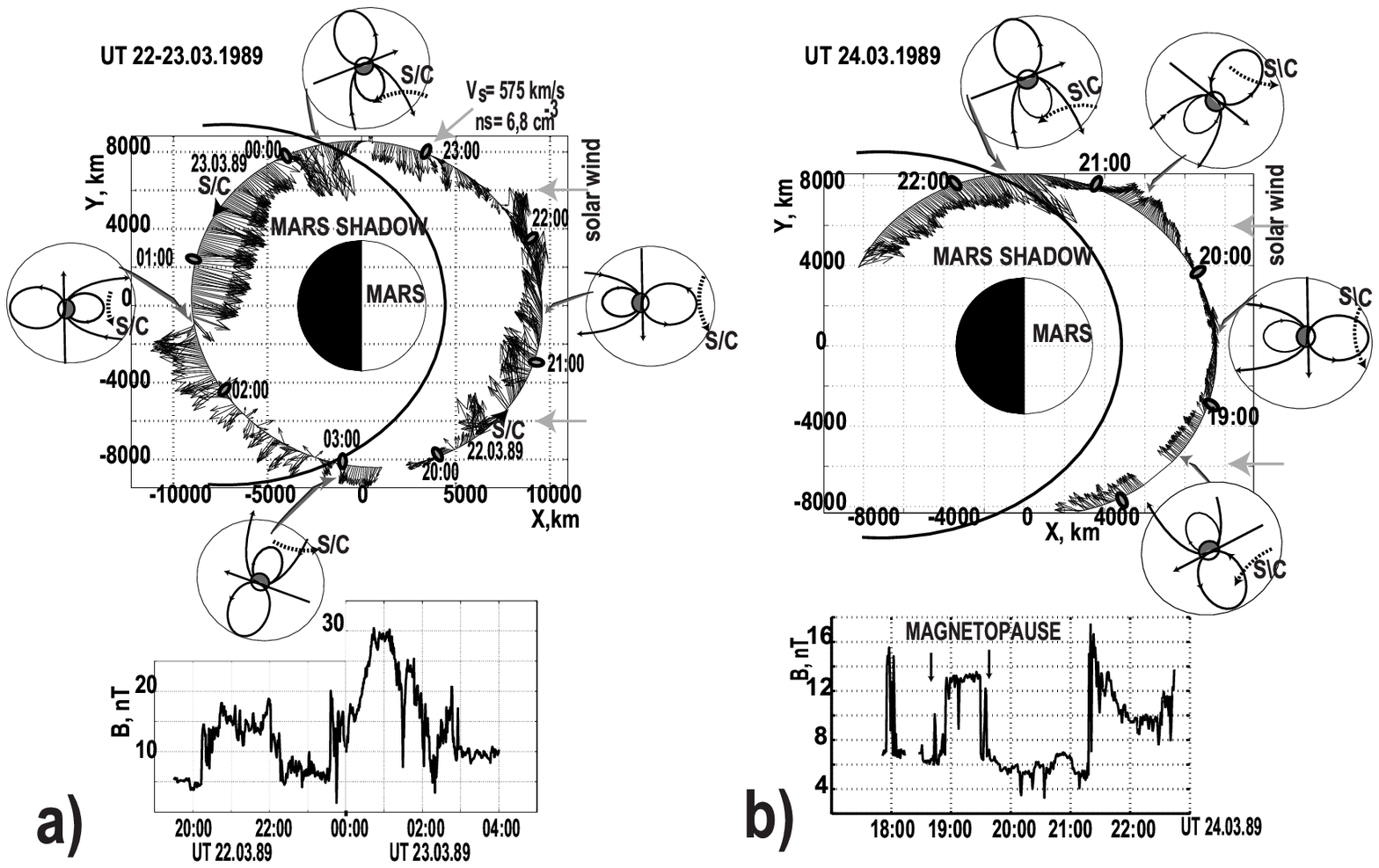}

\vskip 10mm
\begin{minipage}{90mm}
\psfig {file=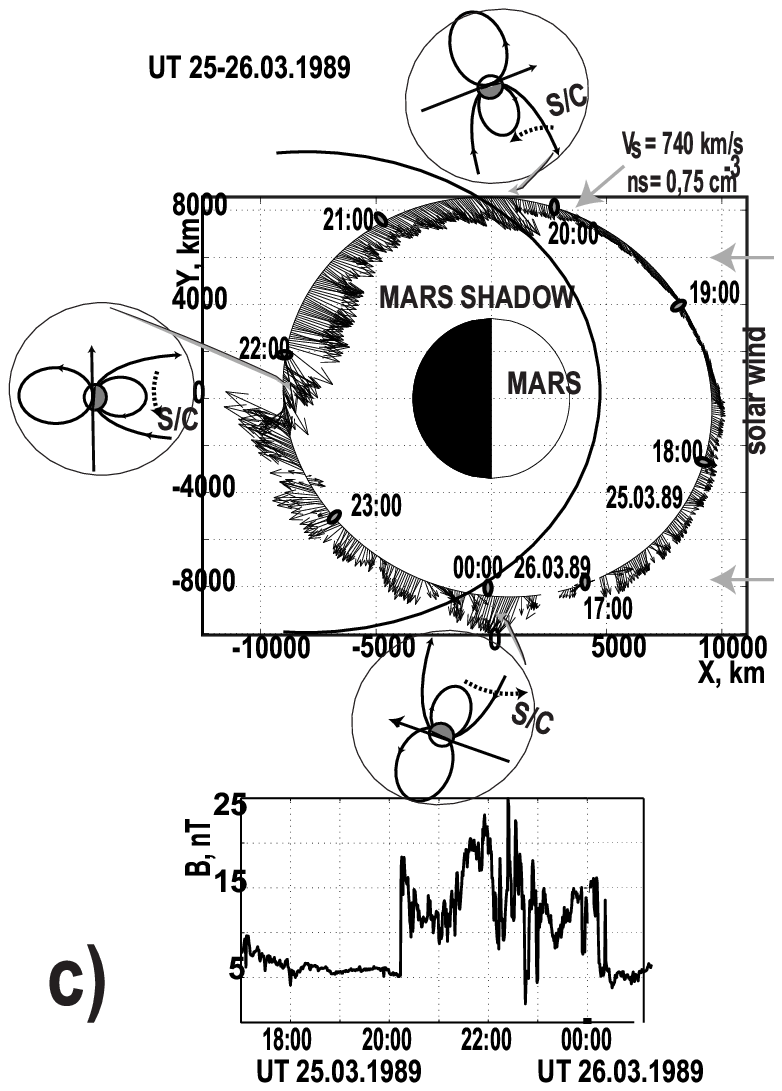}
\end{minipage}
\begin{minipage}{82mm}
\rightskip -15mm
{\small {\bf Fig. 3.} {\bf T}op parts show the orientation and the magnitude of the magnetic field projection onto the Mars ecliptic plane.  {\bf T}he magnitudes of the observed magnetic field versus the time are shown in the bottom parts.
{\bf T}he expected directions of the
Phobos magnetic field are depicted by the schematic sketch of Phobos
magnetosphere. {\bf a)} The data from 20:00 on March 22, 1989 to 04:00 on March 23, 1989. {\bf b)} The data from 17:45 to 22:45 on March 24, 1989.
{\bf c)} The data from 17:00 on March 25, 1989 to 00:45 on March 26, 1989.}
\vskip 8mm
The
magnitude of the observed magnetic field versus the time is
represented in the bottom panels of Fig.~3. The velocity V$_s$ and
density n$_{s}$ of the solar wind are also given (the
units for velocity and density are km/s and cm$^{-3}$, respectively).
To compare the model data with the measurements, we have chosen the
appropriate orientation of the Phobos magnetic moment and depicted the
schematic models of Phobos magnetosphere in Fig.~3. The
corresponding positions of the spacecraft are indicated by the dotted
lines. The big grey arrows going from the schemes indicate the
corresponding place of the measurements. It is easily seen that the
directions of the model magnetic field coincide with that of the
measured one. When comparing the fields, one should take into
consideration that there exist the differences between the measurements
of the magnetic field of the Phobos vicinity in the Mars wake and that in
the solar wind.
\end{minipage}

Figure~3b shows a unique case, when the dynamic pressure of the solar
wind seems dropped and the manifestation of the Phobos magnetic
field can be detected over the entire orbit. At least, the rotation of
the magnetic field direction is clearly observed. This fact gives the
additional evidence that the magnetic field observed during
18:43--19:41 on March 24, 1989 can be associated with an intrinsic
magnetic field of Phobos.

\section*{About the Measurements in the Martian Wake}

It is more easy to distinguish the Phobos magnetic field signature in
the Martian wake rather than in the solar wind. The Martian wake is the
most appropriate region for testing the presence of the Phobos magnetic
field and the most disputable one. The rotation of the field direction
in agreement with the manifestation of the Phobos magnetic field in the
Martian wake is demonstrated in Fig.~3.

To compare the magnetic
signature obtained near corotating region of Phobos (Fig.~3) with
one obtained in the ``pure'' wake of Mars, in Fig.~4 we present
the magnetic field measurements, which were acquired by the spacecraft
when its orbit did not coincide with that of Phobos.

The orientation of the field in the Martian wake is governed by the
direction of the solar wind magnetic field (Russell et al., 1995).
Figure~4a shows a typical example of the magnetic signatures when the
magnetic field of the solar wind points away from the Sun. The data
shown in Fig.~3 are entirely different from that in
Fig.~4a.  Another example illustrating the magnetic signatures, when the
magnetic field of the solar wind points toward the Sun, is shown in
Fig.~4b. The change in polarity of magnetic field components in the
Martian wake can be similar in Fig.~3 and 4b. In Figure~3,
however, in the vicinity of terminator the magnetic field features,
which are caused by the corotating region near Phobos, and the absence
of filamentation of the magnetic field are typical manifestations of
the Phobos magnetic field in the Mars wake.

Under the favorable conditions depending on the relative position of
Phobos and spacecraft, the manifestation of the Phobos magnetic field
was observed and the expected directions of the magnetic field is
always similar to those depicted in Fig.~3 when the spacecraft was
in the vicinity of Phobos within the Mars wake.
\vskip 10mm
\psfig {file=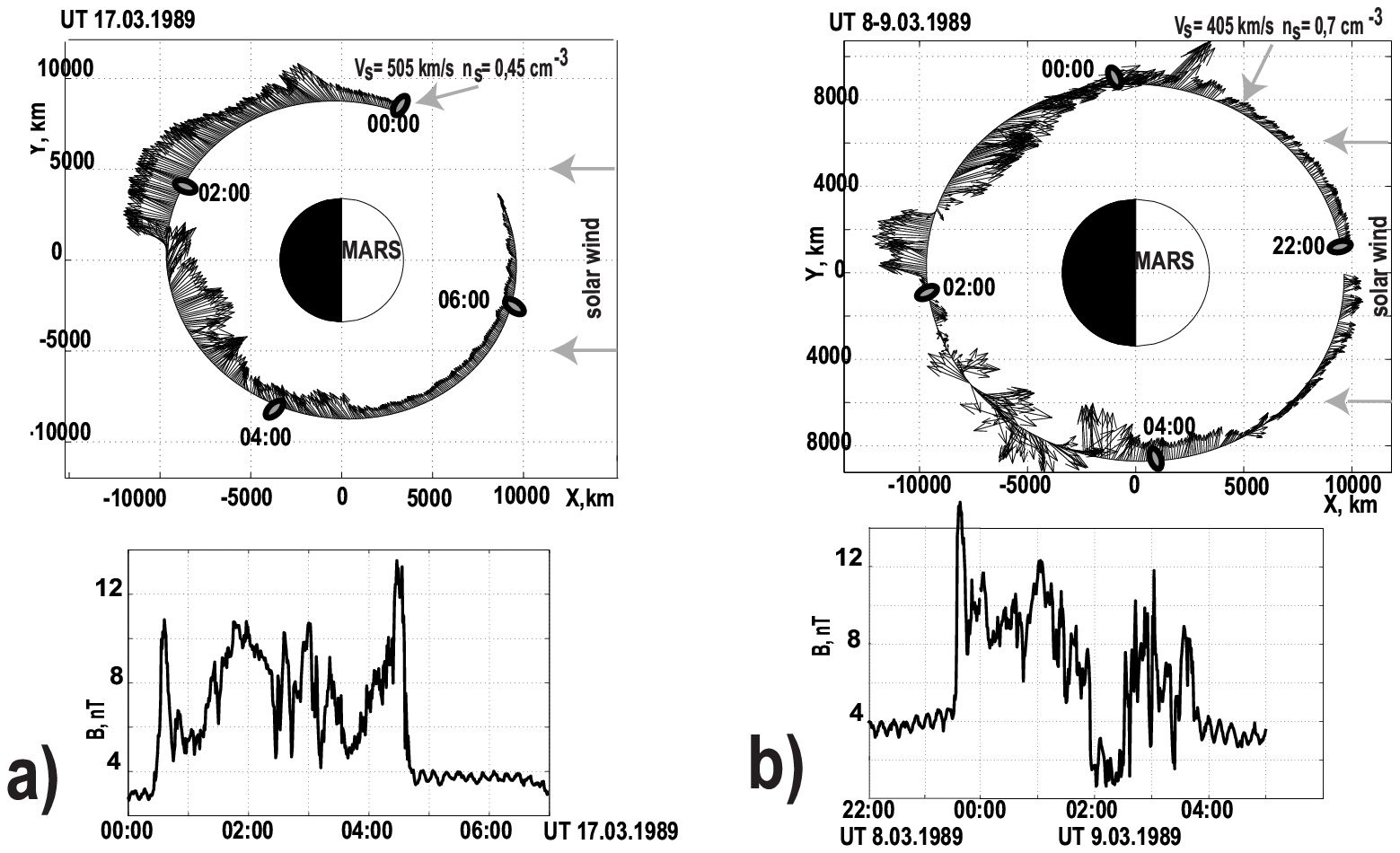}
\vskip 10mm
{\small {\bf Fig. 4.} {\bf T}he same as in Fig. 3, but {\bf a)} - the data from 00:00 to 06:45 on March 17, 1989; {\bf b)} - the data from 22:00 on March 8 to 05:45 on March 9, 1989.}

\section{Conclusion}
The trajectory of {\it Phobos-2} provided the collection of data in the
regions that are relevant for the investigation of Phobos vicinity and
have not been explored before. In this study we considered the
evidences of the existence of Phobos magnetic field from the {\it
Phobos-2} data acquired on 22--24 March, 1989. We could directly probe
regions to observe the pure planetary magnetic field of Phobos during
the closest approaches of the spacecraft to the Phobos surface at the
distance of 170--180 km.

Source with an equivalent magnetic moment $\simeq10^{15}$ A$\cdot$m$^2$
within Phobos leads to the development of an obstacle for solar wind
flow around Phobos with the subsolar stand-off distance of the
deflection about 16--17 Phobos radii. In the plane coinciding with the
projection onto the ecliptic plane of Mars the magnetic moment of
Phobos points to the trailing side. In the projection onto the
$Z$-axis of the coordinate system used, the magnetic moment of Phobos
points out in the negative direction of $Z$-axis. In the present paper,
the direction of the Phobos magnetic moment is determined at a
qualitative level. To obtain its direction more precisely, special care
should be taken when calculating the orbits. The magnetization of the
Phobos substance is 0.15 CGS. It should be noted that there are some
meteorites with a magnetization of 3 CGS (Gus'kova, 1972).

Our results confirm the existence of a corotating part of the measured
magnetic fields. Such a rotation of the magnetic field direction near
Phobos gives an additional evidence that an intrinsic magnetic field
of Phobos does exist.

In conclusion, it is necessary to point out that solving the problems
mentioned in this paper, in the context of the ``Celestial Mechanics''
experiment, is only at the very beginning.

\section*{ACKNOWLEDGEMENTS}
We thank Dr. G.~Kotova for providing some unpublished plasma data
(experiment TAUS).

\end{document}